\documentclass[aps,prb,amsmath,amssymb,reprint,superscriptaddress,preprintnumbers,showpacs,intlimits]{revtex4-1}
\pdfoutput=1
\usepackage{bm,latexsym,mathrsfs,enumerate,color}
\usepackage[mathcal]{euscript}
\usepackage{hyperref}
\usepackage{graphicx}
\renewcommand{\vec}[1]{\bm{#1}}
\begin{document}
%\preprint{\textcolor[rgb]{0.00,0.50,0.75}{{\texttt{Draft v3.1}}}}

\title{Out-of-surface vortices in spherical shells}

\author{Volodymyr P. Kravchuk}
 \email{vkravchuk@bitp.kiev.ua}
 \affiliation{Bogolyubov Institute for Theoretical Physics, 03143 Kiev, Ukraine}
\author{Denis D. Sheka}
%     \email{sheka@univ.net.ua}
\affiliation{Taras Shevchenko National University of Kiev, 01601 Kiev, Ukraine}

\author{Robert Streubel}
% \email{r.streubel@ifw-dresden.de}
 \affiliation{Institute for Integrative Nanosciences, IFW Dresden, 01069 Dresden, Germany}
\affiliation{Material Systems for Nanoelectronics, Chemnitz University of Technology, 09107 Chemnitz, Germany}

\author{Denys Makarov}
% \email{d.makarov@ifw-dresden.de}
 \affiliation{Institute for Integrative Nanosciences, IFW Dresden, 01069 Dresden, Germany}

\author{Oliver G. Schmidt}
% \email{o.schmidt@ifw-dresden.de}
 \affiliation{Institute for Integrative Nanosciences, IFW Dresden, 01069 Dresden, Germany}
\affiliation{Material Systems for Nanoelectronics, Chemnitz University of Technology, 09107 Chemnitz, Germany}

\author{Yuri Gaididei}
% \email{ybg@bitp.kiev.ua}
 \affiliation{Bogolyubov Institute for Theoretical Physics, 03143 Kiev, Ukraine}

\date{\today}

%%%%%%%%%%%%%%%%%%%%%%%%%%%%%%%%%%%%%%%%%%%%%%%%%%%%%%%%%%%%%%%%%%%%%70
%
%         ABSTRACT
%
%%%%%%%%%%%%%%%%%%%%%%%%%%%%%%%%%%%%%%%%%%%%%%%%%%%%%%%%%%%%%%%%%%%%%70

\begin{abstract}
The interplay of topological defects with curvature is studied for out-of-surface magnetic vortices in thin spherical nanoshells. In the case of easy-surface Heisenberg magnet it is shown that the curvature of the underlying surface leads to a coupling between the localized out-of-surface component of the vortex with its delocalized in-surface structure, \emph{i.e. polarity-chirality coupling}.
\end{abstract}


% 75.10.Hk    Classical spin models
% 75.40.Mg    Numerical simulation studies
% 05.45.-a    Nonlinear dynamics and nonlinear dynamical systems.
% 72.25.Ba    Spin polarized transport in metals
% 85.75.-d    Magnetoelectronics; spintronics: devices exploiting spin polarized transport or integrated magnetic fields

%%%%%%%%%%%%%%%%%%%%%%%%%%%%%%%%%%%%%%%%%%%%%%%%%%%%%%%%%%%%%%%%%%%%%70

\maketitle
Understanding of the interplay between  geometry and topology of condensed matter order  is of fundamental importance in several physical and biophysical contexts, and this combination raises a number of unsolved questions. Examples are thin layers of superfluids and superconductors\cite{Tempere09,Fomin12}, nematic liquid crystal shells\cite{Lopez-Leon11}, viral shells\cite{Ganser99}, and cell membranes\cite{McMahon05}. A considerable effort has been invested in understanding the role  of the coupling between in-surface order and curvature of the underlying surface.\cite{Bowick09,Turner10} The topological defects  of 2D in--surface vector fields are characterized  by a winding number for the phase--like variable: vorticity  $q\in\mathbb{Z}$ (topological charge of $\pi_1(S^1)$ homotopy group). On curved surfaces the Gaussian curvature leads to screening topological charges\cite{Vitelli04}. Vortices in curved superfluid films are a typical example of such kind of defects.

Vortices in magnets belong to a more general type of topological defects.
In addition to the  vorticity, the magnetic vortex is also characterized by the polarity $p=\pm1$, which describes the vortex core magnetization. The topological properties of magnetic vortices are characterized by the  relative homotopy group $\pi_2(S^2, S^1)$ \cite{Volovik03} and depend on both vorticity and polarity.
Magnetic vortices were intensively studied during last decades for the sake of applications in nanomagnetism as high--density magnetic storage devices \cite{Zhu00} and miniature sensors \cite{Guimaraes09}. Investigations of different aspects of magnetic vortex statics and dynamics were mainly restricted to flat structures. In such nanomagnets, the vortex appears as a ground state in sub--micrometer sized magnets due to competition between short--range exchange interaction and long--range dipole interaction \cite{Hubert98,Guimaraes09}. The ground state of smaller samples is typically characterized by in--plane quasi--uniform magnetization. Contrary to in--surface, a quasi--uniform magnetization distribution in thin spherical shells is forbidden for topological reasons; Instead, two oppositely disposed vortices are expected.

In a flat nanomagnet, the vortex state is degenerated with respect to  polarity. Hence, one can link the vortex polarity to the bit of information with possible spintronics applications \cite{Nakano11,*Yu11}. One of the consequence of a more complicated topology of the magnetic vortex is a gyroscopical force that depends on both vorticity and polarity of the vortex. Therefore, the vortex polarity can be switched by exciting the gyroscopical motion. The switching thresholds for the two polarities are only equal in ideally flat structures \cite{Curcic08a}. Experiments on permalloy platelets have revealed a relatively large asymmetry in thresholds  \cite{Chou07,Curcic08a}, which originates from the lack of the mirror--symmetry of rough thin--film structures \cite{Vansteenkiste09a}. This indicates the interplay between the vortex polarity and the curvature of the underlying surface.

The influence of a curvature on magnetic properties have been studied both experimentally and theoretically for geometries of cylinder \cite{Saxena98,*Landeros}, torus \cite{Carvalho-Santos08,*Carvalho-Santos10}, cone \cite{Freitas05,*Moura07a} and hemispherical cap structures \cite{Albrecht05,*Ulbrich06,*Sapozhnikov12}. In this respect, we recently demonstrated experimentally by means of $x$-ray magnetic circular dichroism photoemission electron microscopy (XMCD-PEEM) the stability of magnetic vortex in thin permalloy (Ni$_{80}$Fe$_{20}$) films on spherical particles \cite{Streubel12}. A precise theoretical description of peculiarities of vortices on spherical surfaces is not available in literature. Most theoretical studies are limited to skyrmion--like solutions \cite{Dandoloff11}.

In this paper, we study the structure of magnetic vortices on a thin spherical shell with an easy--surface anisotropy. Using anisotropic Heisenberg model, we find possible solutions of the vortex type. On the contrary to vortices in flat magnets, there is an interplay between the localized out--of--surface and the delocalized in-–surface structure. In other words, the vortex core plays the role of a charge source for the vortex phase structure.

The magnetic energy of a classical Heisenberg easy--surface ferromagnet has the following form:
\begin{equation} \label{eq:E-gen}
E = A \int\mathrm{d}\vec r \left[ - \vec m\cdot\vec{\nabla}^2 \vec m +  \frac{\left(\vec{m}\cdot \vec{n}\right)^2}{\ell^2} \right],
\end{equation}
with the exchange constant $A$, the anisotropy constant $K>0$, magnetic length $\ell = \sqrt{A/K}$, 
%\footnote{The anisotropy is supposed to be sufficiently strong to overcome dipolar field contribution.}
the surface normal $\vec{n}$, and the integration is over volume of the spherical shell. In the following, we use the local spherical reference frame for the unit magnetization vector $\vec{m} = (m_r, m_\vartheta, m_\chi) = (\cos\Theta, \sin\Theta\cos\Phi, \sin\Theta\sin\Phi)$. Here, angular magnetic variables $\Theta=\Theta(\vec r)$ and $\Phi=\Phi(\vec r)$ describe the magnetization distribution with respect to the spherical coordinates $(r,\vartheta,\chi)$ of the radius--vector $\vec{r}$. Hereinafter we consider a case of thin and high anisotropy shell: $h\ll\ell\ll L$, where $h$ is thickness of the shell and $L$ is its inner radius. Therefore we assume that the magnetization does not depend on the radial coordinate $r$. On the indicated conditions the total magnetic energy in terms of the local reference frame reads
%\begin{subequations} \label{eq:Energy}
%\begin{equation} \label{eq:E0}
%E_0 = \frac{A}{2} \int \mathrm{d}\vec r \left[(\nabla\Theta)^2+\sin^2\Theta (\nabla\Phi)^2 + \frac{\cos^2\Theta}{\ell^2}\right]
%\end{equation}
%and an additional energy term originating from the curvature
%\begin{equation} \label{eq:E-crv}
%\begin{split}
%&E_{\text{crv}} = A \int \frac{\mathrm{d}\vec r}{r^2}
%\Biggl[ 1+\sin^2\Theta\frac{\cos2\vartheta}{2\sin^2\vartheta} \\
%& +\cos\Phi\partial_{\vartheta} \Theta - \sin\Theta\cos\Theta \sin\Phi \partial_{\vartheta} \Phi  +\! \frac{\sin\Phi}{\sin\vartheta} \partial_{\chi } \Theta\! \\
%&+\!\left( \cos\Theta \cos\Phi\! +\! \cot\vartheta\sin\Theta \right)\sin\Theta\frac{\partial_{\chi}\Phi}{\sin\vartheta}\Bigr]\!.
%\end{split}
%\end{equation}
%\end{subequations}
\begin{equation}\label{eq:Energy}
\begin{split}
E&=Ah\int\limits_0^{2\pi}\mathrm{d}\chi\int\limits_0^\pi\mathrm{d}\vartheta\sin\vartheta\Biggl\{(\partial_\vartheta\Theta+\cos\Phi)^2\\
&+\frac{1}{\sin^2\vartheta}(\partial_\chi\Theta+\sin\vartheta\sin\Phi)^2+\frac{L^2}{\ell^2}\cos^2\Theta\\
&+\sin^2\Theta\Bigl[(\partial_\vartheta\Phi-\sin\Phi\cot\Theta)^2\\
&+\frac{1}{\sin^2\vartheta}(\partial_\chi\Phi+\cos\vartheta+\sin\vartheta\cot\Theta\cos\Phi)^2\Bigr]\Biggr\}
\end{split}
\end{equation}
%Firstly, we discuss solutions of the general form $\Theta=\Theta(\vartheta)$, $\Phi=\Phi(\vartheta)$ for infinitesimally thin sphere with the radius $L$.
In this case, the static magnetization configuration the energy functional \eqref{eq:Energy} produces the Euler-Lagrange equations
\begin{subequations} \label{eq:Theta-Phi-equaions}
\begin{align}
\label{eq:Theta}
&\nabla^2\Theta - \sin\Theta\cos\Theta \left[({\nabla}\Phi)^2-1+\cot^2\vartheta-\frac{L^2}{\ell^2}\right] \\
&=2\frac{\sin^2\Theta}{\sin\vartheta}\left[\Xi\partial_\chi\Phi - \partial_\vartheta( \sin\vartheta \cos\Phi) \right],\nonumber\\
\label{eq:Phi}
& \nabla\!\cdot\!\!\left(\sin^2\Theta{\nabla}\Phi\right) = -2\frac{\sin^2\Theta}{\sin\vartheta}[\Xi\partial_\chi\Theta +\sin\vartheta\sin\Phi\partial_\vartheta\Theta],
\end{align}
\end{subequations}
where $\Xi\equiv\cot\Theta\cot\vartheta-\cos\Phi$ and $\nabla$-operators denotes the angular parts of the corresponding differential operators in the spherical local basis.
%$\vec{\nabla} \equiv \vec{e}_\vartheta\partial_\vartheta + \vec{e}_\chi\csc{\vartheta} \partial_\chi$.

In the case of a high easy--surface anisotropy $(\ell\rightarrow0)$ the solution of Eqs.~\eqref{eq:Theta-Phi-equaions} which minimize the energy \eqref{eq:Energy} reads $\Theta=\pi/2$, $\Phi=\mathrm{const}$. This is a vortex solution where the magnetization is confined within the sphere surface except of two diametrically opposite point singularities -- vortex cores. Such kind of ``in-surface" vortices is well studied in different media\cite{Bowick09,Turner10}. Here we demonstrate that taking into account the out-of-surface structure of the vortex core (finite $\ell>0$) essentially changes the vortex state properties in case of the curved surface as compared with planar magnets. In the following we consider only the azimuthally symmetric vortex solution $\Theta=\Theta(\vartheta)$, $\Phi=\Phi(\vartheta)$ by analogy with the planar vortices.

The out-of-surface magnetization of the vortex core, so--called polarity, takes two values, $p=\pm1$ (outward and inward). The magnetization distribution can be analyzed asymptotically near the vortex center ($\vartheta=0$), see Appendix~\ref{app:asympt}. The size of the vortex core is determined by the vortex out--of--surface magnetization, $\cos\Theta\approx p(1-\vartheta^2/2\vartheta_c^2)$, with $\vartheta_c\ll1$. Moreover, the in--surface magnetization is described by the angular variable $\Phi\approx\Phi_0-p \sin\Phi_0\vartheta^2/(4\vartheta_c)$, where the constant $\Phi_0$ will be determined later. Although the asymptotic limit of out--of--surface component is similar to that of a vortex in planar magnets \cite{Feldtkeller65}, the phase $\Phi$ depends on the vortex polarity $p$, which is a distinct feature compared with the constant value in planar vortices. Since the out--of--surface magnetization has an exponentially localized structure, the following Ansatz function (similar to the vortices in planar magnets \cite{Feldtkeller65}) can be used for description:
\begin{equation} \label{eq:cores-tst}
\cos\Theta=p_1e^{-\frac12\left(\frac{\vartheta}{\vartheta_c}\right)^2}+ p_2e^{-\frac12\left(\frac{\pi-\vartheta}{\vartheta_c}\right)^2},
\end{equation}
where $p_1$ and $p_2$ are polarities of the vortices at poles. Now, we consider the vortex in--surface magnetization. Accurate within the vortex core corrections, $\Phi$--distribution can be described by the following equation:
\begin{equation} \label{eq:Phi-g}
\partial_{uu}\Phi=-g(u)\sin\Phi, \quad g(u) = -\frac{4e^u\partial_u m_r(u)}{1+e^{2u} },
\end{equation}
with $u=\ln\tan(\vartheta/2)$ and $m_r(u) = \cos\Theta$. The function $g(u)$ consists of two peaks localized near the vortex cores at $u_c\approx \ln\cot(\vartheta_c/2)$.
%Symmetry of the function $g(u)$ is determined by the relative polarities of the vortices.

%%==================================================================\
%\begin{figure}
%\includegraphics[width=\columnwidth]{g_function}
%\caption{Function $g(u)$ from Eq.~\eqref{eq:Phi-g} for $\vartheta_c=0.05$.
%%Left plot corresponds to the case of the same polarities of the vortices in poles $\vartheta=0$ and $\vartheta=\pi$ and the right plot corresponds to the case of opposite polarities.
%}
%\label{fig:g-function}
%\end{figure}
%%==================================================================/

In order to analyze the $\Phi$--distribution outside cores, we use the stepwise vortex shape model for the out--of--surface magnetization, $m_r(u) \approx p_1 - p_1 h(u+u_c) + p_2 h(u-u_c)$ with the Heaviside step function $h(u)$. Using this approach, $g(u)$ becomes $g(u)\approx\frac{\pi}{2}\vartheta_c[p_1\delta(u+u_c)-p_2\delta(u-u_c)]$.
%, where the coefficient $\frac{\pi}{2}\vartheta_c$ provides an approximate equality of area under modeled and real peaks.
The consequence of such a model is that the localized out--of--surface structure plays the role of the charge density for the delocalized in--surface structure. The solution of this model, which satisfies the Neumann boundary conditions $\partial_u\Phi(\pm\infty)=0$, has the implicit form:
\begin{align} \label{eq:Phi-piece-wise}
&\Phi(u) = \Phi_0 - \frac{\pi}{2} \vartheta_c p_1\sin\Phi(-u_c)\left[ (u+u_c)_+ -(u-u_c)_+\right],\nonumber\\
&p_1\sin\Phi(-u_c) = p_2\sin\Phi(u_c),
\end{align}
with $u_+\equiv u h(u)$. The further analysis essentially depends on the relative orientations of vortices.

%==================================================================\
\begin{figure}
\includegraphics[width=\columnwidth]{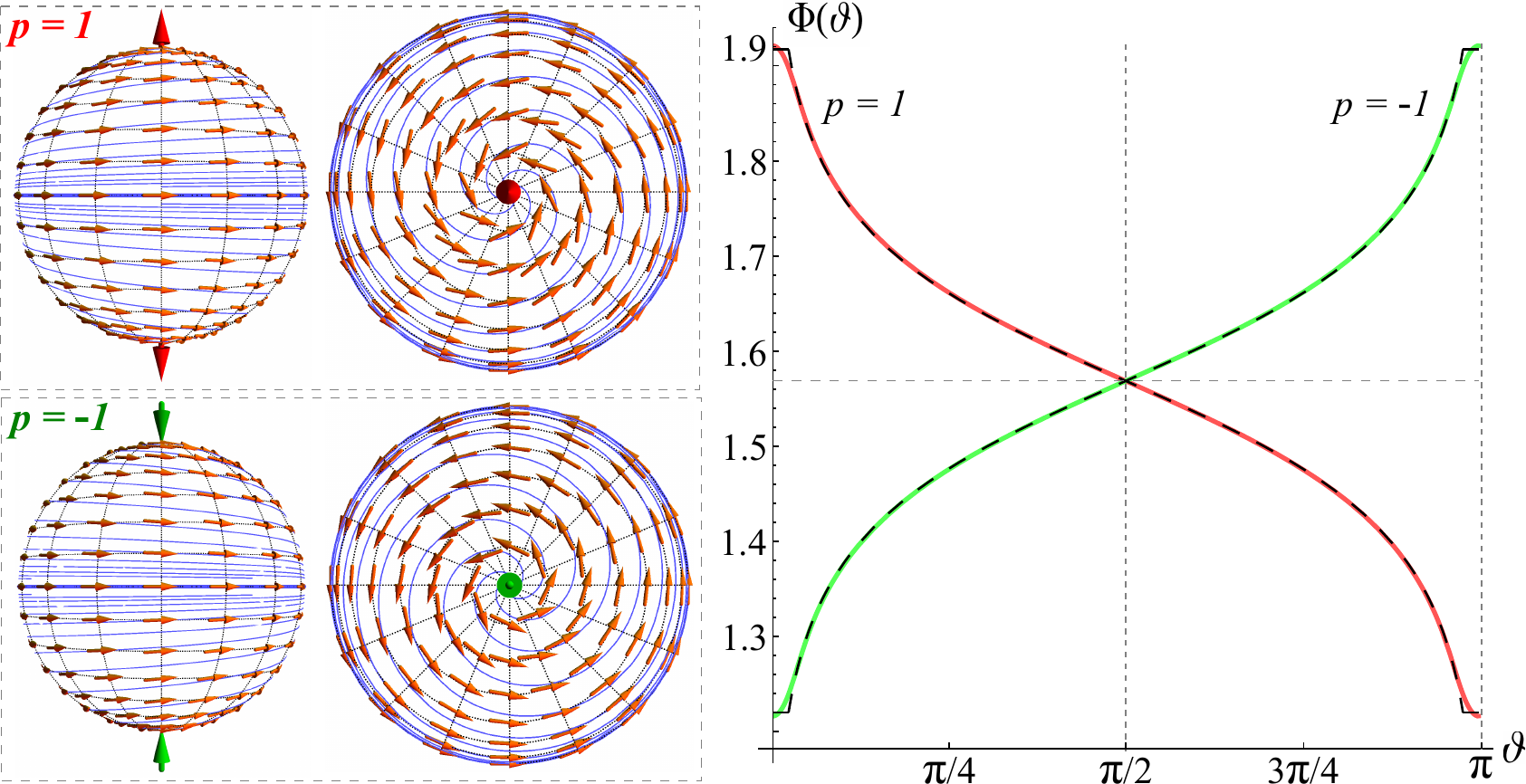}
\caption{Possible vortex phases $\Phi$ for the case of same polarities $p_1=p_2=p$ (right inset). Solid lines correspond to the exact numerical solution of Eq.~\eqref{eq:Phi}, where the out-of-surface component is chosen in form of Eq.~\eqref{eq:cores-tst} with $\vartheta_c=0.05$. The corresponding approximate solutions \eqref{eq:Phi-sol-appr} are indicated by dashed lines. The corresponding magnetization distribution on the sphere surface is schematically shown in the left inset using arrows and stream lines.}
\label{fig:Phi-num}
\end{figure}
%==================================================================/

For the \textit{case of same polarities} $(p_1=p_2=p)$, the solution $\Phi$ that minimizes the energy \eqref{eq:Energy} takes the following explicit form outside vortex cores ($\vartheta_c\ll1$):
\begin{equation} \label{eq:Phi-sol-appr}
\Phi(\vartheta) \approx \pm\frac{\pi}{2}\left(1-p\vartheta_c \alpha\ln\tan\frac{\vartheta}{2}\right),
\end{equation}
where $\alpha$ is solution of the equation $\alpha=\cos(\alpha\vartheta_cu_c\pi/2)$. For details see Appendix~\ref{app:sol}. Since $\vartheta_cu_c\ll1$, so $\alpha\lessapprox1$.
Accordingly, $\Phi$ takes constant values inside the vortex cores, in particular, $\Phi(\vartheta<\vartheta_c) \approx \Phi_0  =\pm\pi/2(1-p\vartheta_c\ln\vartheta_c)$.  Energy of the vortex state with exception of the core energy $E_c$ depends on the core size, $E-E_c\propto \vartheta_c^2 u_c$.
%==================================================================\
\begin{figure}
\includegraphics[width=0.8\columnwidth]{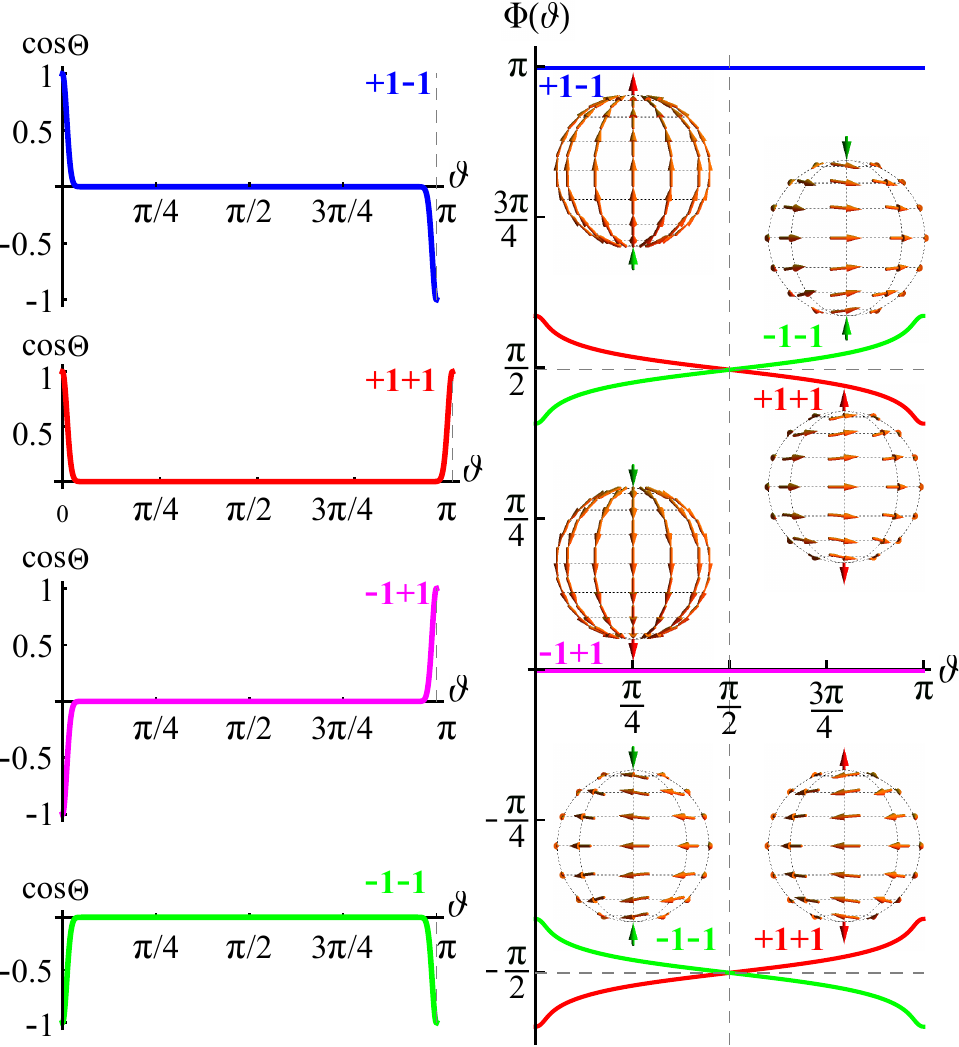}
\caption{All possible vortex states of the spherical surface. The left column demonstrates the model out-of-surface magnetization distribution given by Eq.~\eqref{eq:cores-tst} for all possible combinations of polarities. The right graph shows the corresponding distributions of the phase $\Phi(\vartheta)$. For the correspondence, we use the notation $p_1p_2$, e.g. "+1-1" means $p_1=+1$ and $p_2=-1$. }
\label{fig:all_vortices}
\end{figure}
%==================================================================/
The dependence $\Phi(\vartheta)$ is indicated in Fig.~\ref{fig:Phi-num} by dashed lines. The approximate solution is in a good agreement with the numerical solution of Eq.~\eqref{eq:Phi}, where the out-of-surface component was chosen according to Eq.~\eqref{eq:cores-tst}. It should be emphasized that the phase of the vortex on a spherical surface gains a coordinate dependence given by Eq.~\eqref{eq:Phi-sol-appr} and has the maximum amplitude in the center of each vortex, as opposed to the planar vortex.

%==================================================================\
\begin{figure}
\includegraphics[width=\columnwidth]{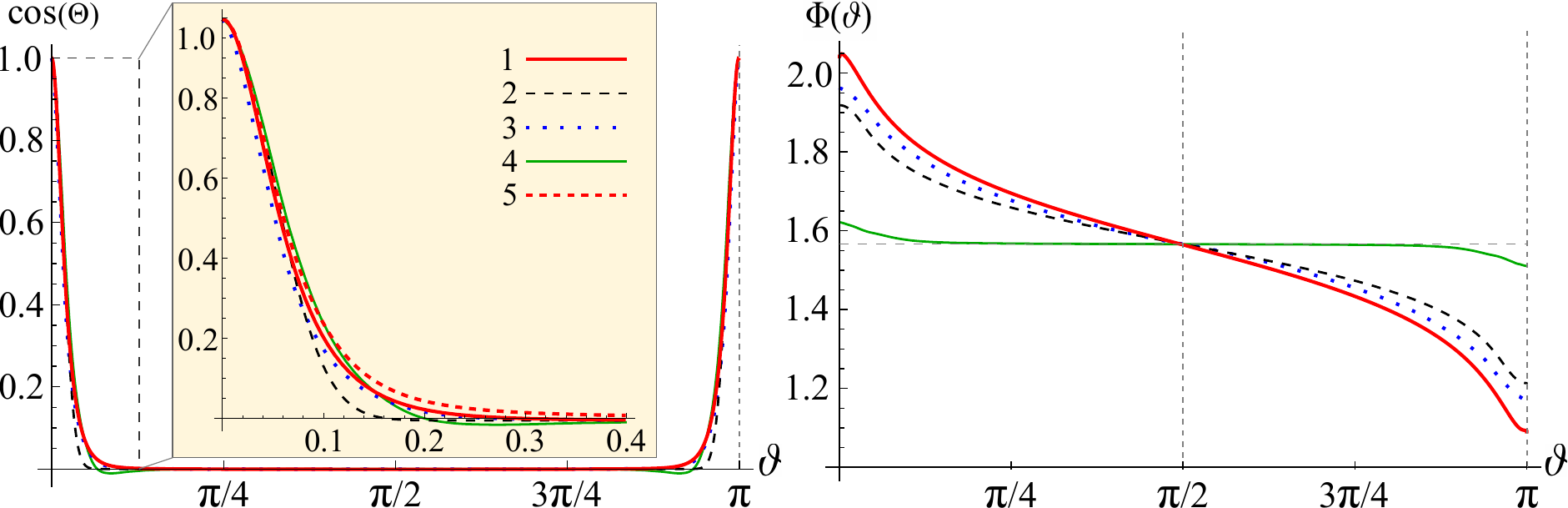}
\caption{Structure of the vortex state for the case of same polarities ($p_1=p_2=1$) obtained by different methods. Line~1 --- exact numerical solution of Eq.~\eqref{eq:Theta-Phi-equaions} with boundary conditions $\Theta(0)=\Theta(\pi)=0$ and $\Phi'(0)=\Phi'(\pi)=0$ and $\ell/L=0.05$. Line~2 --- (left) the Ansatz~\eqref{eq:cores-tst} and (right) corresponding solution of Eq.~\eqref{eq:Phi} where the function $\Theta(\vartheta)$ is determined by Eq.~\eqref{eq:cores-tst} with an angular vortex core size $\vartheta_c=\ell/L=0.05$. This line coincides with line "$p=1$" in Fig.~\ref{fig:Phi-num}. Line~3 and line~4 correspond to micromagnetic simulations of types (i) and (ii), respectively. (see text for details). Line~5 in the left graph shows the out-of-surface structure of the vortex core for the case of opposite polarities, when $p_1=-p_2=1$ and $\Phi\equiv\pi$.}
\label{fig:comparison}
\end{figure}
%==================================================================/

For the \textit{case of opposite polarities} $(p_1=-p_2=p)$, the energy reaches its minimum for the trivial solutions $\Phi=\pi$ for $p=1$ and $\Phi=0$ for $p=-1$ (as earlier, we consider the case $\vartheta_c\ll1$), see Appendix~\ref{app:sol} for details. Such a solution can be considered as a three dimensional generalization of well-known onion state in narrow nanorings. We refer to this solution as the pumpkin state. The energy of the pumpkin state, $E-E_c\propto - \vartheta_c$, is lower than for the vortex state. Namely, the energy gain is $\Delta E\propto \vartheta_c$. Nevertheless, it should be emphasized that these two states are separated by a high energy barrier related to the polarity switching of one of the vortices. Therefore, we suggest that both states can be realized experimentally even at room temperature. All possible vortex--like states described above are presented in Fig.~\ref{fig:all_vortices}.

In this paragraph, a comparison between the obtained analytical results and the exact numerical solutions as well as the micromagnetic simulations is given. The numerical solution of Eq.~\eqref{eq:Theta-Phi-equaions} with the boundary conditions $\Theta(0)=\Theta(\pi)=0$ and $\Phi'(0)=\Phi'(\pi)=0$ leads to a vortex structure for the case of same polarities $p_1=p_2=1$. The obtained out-of-surface structure of the vortex core $\cos\Theta$ is quite close to the model solution \eqref{eq:cores-tst} when $\vartheta_c=\ell/L$ (left graph of Fig.~\ref{fig:comparison}). The solution for the vortex phase $\Phi(\vartheta)$ for same boundary conditions corresponds to two-fold degenerated solutions of the type given by Eq.~\eqref{eq:Phi-sol-appr} with opposite chiralities. One of such solutions is plotted in the right graph (Fig.~\ref{fig:comparison}). The exact solution $\Phi(\vartheta)$ has slightly larger amplitude of turning compared with the model solution. This originates from a larger effective core size $\vartheta_c$ of the exact solution compared with the model given in Eq.~ \eqref{eq:cores-tst} (left graph of Fig.~\ref{fig:comparison}).

In order to verify our results, we performed two types of micromagnetic simulations by using the \textsf{OOMMF} code \cite{OOMMF}. A thin spherical shell was simulated considering (i) local magnetic interaction in the form of Eq.~\eqref{eq:E-gen} and (ii) exchange and non-local magnetostatic interaction. In both cases, the material parameters \footnote{\label{fnote:params}The spherical shell with an inner radius $L=150\,nm$ and a thickness $h=15\,nm$ was simulated using a unit cell size $\Delta=5\,nm$. For both types of simulations, the exchange constant $A=1.3\times10^{-11}\,J/m$ and the saturation magnetization $M_S=6\times10^{5}\,A/m$ were used. In simulations of type (i) we used an uniaxial anisotropy with a spatially varying hard axis orientated along the radial vector $\vec e_r$ and an anisotropy constant $K=-2.2\times10^5\,J/m^3$. For both types of simulations the characteristic magnetic length were approximately equal: $\sqrt{A/|K|}\approx\sqrt{A/2\pi M_S^2}\approx7.6\,nm$.} were chosen in the way to provide the same ratio of characteristic magnetic length and the sphere radius $\ell/L=0.05$. Physically, these two types of simulations are equivalent in case of vanishing thickness, when the magnetostatic interaction can be reduced to the easy-surface anisotropy.

The simulations of type (i) confirm the analytical results with a high accuracy. In the case of same polarities, the system relaxes to the vortex state with the additional turning described by Eq.~\eqref{eq:Phi-sol-appr} (line 3 in Fig.~\ref{fig:comparison}). For opposite polarities, the system relaxes to the state with $\Phi\equiv\pi$ or $\Phi\equiv0$, as described above.

According to the simulations of type (ii), in case $p_1=p_2$, the magnetostatic interaction attenuates (but does not suppress) the phenomenon of the vortex phase turning (line 4 in Fig.~\ref{fig:comparison}) due to an increased energy of volume magnetostatic charges. In case of $p_1=-p_2$, the shell of mentioned size relaxes to the vortex state with $\Phi\approx\pm\pi/2$ instead of the pumpkin state, which is preferred for spheres of smaller size \footnote{For example, a spherical shell with an inner radius $L=20nm$ and a thickness $h=4nm$.}. To obtain an approximate criterion of the separation between vortex and pumpkin states, the difference of energies of the pumpkin and vortex states are estimated as following: $\Delta E = \Delta E_{ms}+\Delta E_{ex}$, where $\Delta E_{ms}\sim Lh^2$ is energy increase due to volume magnetostatic charges and $\Delta E_{ex}\sim-\vartheta_c\ell^2h$ is the corresponding exchange energy decrease. Thereby, the pumpkin state is energetically preferable when $L^2h<\ell^3$. The detailed study of ground states of soft-magnetic spherical shells goes beyond the scope of this paper and it is subject of a prospective work.

%Detailed study of the influence of magnetostatics on the vortex structure on a spherical surface has to be made in the future works.

In conclusions, we predict novel features of magnetic vortex in thin spherical shell in comparison with the well known vortex in planar easy-plane magnet. We show that the vortex on a spherical surface gains a coordinate and \emph{polarity-dependent} turning of its phase. An interplay between topological properties of the vortex, namely, its polarity, and the curvature of the underlying surface results breaks the degeneration of the phase--like variable $\Phi$ with respect to the rotation by any constant angle $\Phi_0$. This degeneration is known as well as for $\pi_1$--vortices in different media and for $\pi_2$--vortices in flat magnets. It is instructive to note that the angle $\Phi_0$ in magnetic nanodisks determines the vortex chirality \cite{Gaididei08}. Thus one can speak about \emph{polarity--chirality coupling}.

\appendix

\section{Asymptotic of functions $\Theta(\vartheta)$ and $\Phi(\vartheta)$ in neighborhood of the vortex origin}
\label{app:asympt}
We consider here an asymptotic solution of Eqs.~\eqref{eq:Theta-Phi-equaions} in neighborhood of point $\vartheta=0$. The functions $\Theta(\vartheta)$ and $\Phi(\vartheta)$ can be presented in form of the Tailor series
\begin{subequations}\label{eq:expansion}
\begin{align}
\label{eq:theta-expan}&\Theta\approx\pi h(-p)+p\frac{\vartheta}{\vartheta_c}+\sum\limits_{n=2}^Na_n\vartheta^n,\\
&\Phi\approx\Phi_0+\sum\limits_{n=1}^Nb_n\vartheta^n.
\end{align}
\end{subequations}
where $p$ is the vortex polarity and $h(x)$ is the Heaviside function. The expansion \eqref{eq:theta-expan} satisfies the necessary boundary conditions: $\Theta(0)=0$ for $p=+1$ and $\Theta(0)=\pi$ for $p=-1$. To obtain the asymptotic expansion accurate within terms of order $\mathcal{O}(\vartheta^\nu)$ we can restrict ourselves with $N=\nu+m$, where $m=2$ is the order of Eqs.~\eqref{eq:Theta-Phi-equaions}. Then we substitute the series \eqref{eq:expansion} into \eqref{eq:Theta-Phi-equaions}, expand the obtained equations by the small quantity $\vartheta$ and equate the series coefficients of the same order terms until the order $\nu$. The obtained system results a relation of expansion coefficients in \eqref{eq:expansion}. For the case $\nu=2$ the described procedure results
\begin{equation}
\begin{split}
a_2=&b_1=0,\\
%a_3=&-\frac{2+\vartheta_c^2\left(4+3\frac{L^2}{\ell^2}\right)+6\vartheta_c\cos\Phi_0}{24\vartheta_c^3},\\
%b_1=&0,\\
b_2=&-p\frac{\sin\Phi_0}{4\vartheta_c}.
\end{split}
\end{equation}
Thereby we obtain the following asymptotic expansion accurate within terms of order $\mathcal{O}(\vartheta^2)$
\begin{equation}
\begin{split}
&\Theta\approx\pi h(-p)+p\frac{\vartheta}{\vartheta_c},\\
&\Phi\approx\Phi_0-p\frac{\sin\Phi_0}{4\vartheta_c}\vartheta^2.
\end{split}
\end{equation}

\section{Vortex phase solution $\Phi(\vartheta)$ for the stepwise vortex shape model}
\label{app:sol}
%==================================================================\
\begin{figure}
\includegraphics[width=\columnwidth]{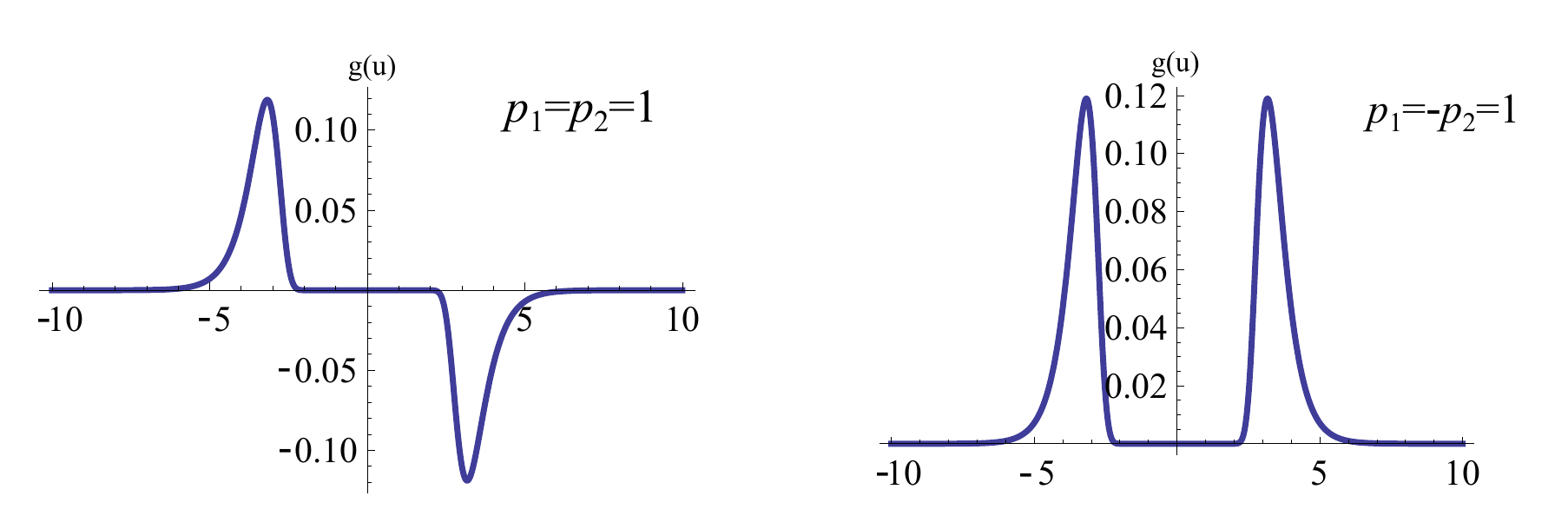}
\caption{Function $g(u)$ from Eq.~\eqref{eq:Phi-g} for $\vartheta_c=0.05$.
%Left plot corresponds to the case of the same polarities of the vortices in poles $\vartheta=0$ and $\vartheta=\pi$ and the right plot corresponds to the case of opposite polarities.
}
\label{fig:g-function}
\end{figure}
%==================================================================/

In the following, we focus on the solution of the Eq.~\eqref{eq:Phi-g}. Using the stepwise vortex shape model $m_r(u) \approx p_1 - p_1 h(u+u_c) + p_2 h(u-u_c)$, one can rewrite \eqref{eq:Phi-g} in the form:
\begin{equation} \label{eq:Phi-g'}
\begin{split}
\partial_{uu}\Phi &= -g(u)\sin\Phi, \\
g(u)&\approx\frac{\pi}{2}\vartheta_c \left[p_1\delta(u+u_c)-p_2\delta(u-u_c) \right],
\end{split}
\end{equation}
where $\delta(u)$ is the Dirac delta function. This approximation agrees well with the real peaked form of $g$-function, see Fig.~\ref{fig:g-function}. The general solution of \eqref{eq:Phi-g'} takes the form:
\begin{equation} \label{eq:Phi-piece-wise-sol}
\begin{split}
\Phi(u) = \Phi_0 +\Phi_1 u - \frac{\pi}{2} \vartheta_c \Bigl[&p_1\sin\Phi(-u_c) (u+u_c)_+ -\\
-&p_2\sin\Phi(u_c)(u-u_c)_+\Bigr]
\end{split}
\end{equation}
with $u_+\equiv u h(u)$ and $h(u)$ being the Heaviside step function. Using the Neumann boundary conditions $\partial_{u}\Phi(\pm\infty)=0$, one can find that $\Phi_1=0$ and
\begin{equation} \label{eq:Phi-piece-wise-1-2}
p_1\sin\Phi(-u_c) = p_2\sin\Phi(u_c).
\end{equation}
Substituting \eqref{eq:Phi-piece-wise-1-2} into \eqref{eq:Phi-piece-wise-sol} results the Eq.~\eqref{eq:Phi-piece-wise}
\begin{equation}
\Phi(u) = \Phi_0 - \frac{\pi}{2} \vartheta_c p_1\sin\Phi(-u_c)\left[ (u+u_c)_+ -(u-u_c)_+\right].
\end{equation}
Thereby for the interval $-u_c<u<u_c$ one can write $\Phi(u)=au+b$, where constants $a$ and $b$ can be found from the system
\begin{equation} \label{eq:Phi-piece-wise-ab}
\begin{split}
&p_1\sin(-au_c+b) = p_2\sin(au_c+b),\\
&a=-\frac{\pi}{2}\vartheta_cp_1\sin(-au_c+b).
\end{split}
\end{equation}
The further analysis essentially depends on the relative polarities of vortices. For the case of equal polarities $p_1=p_2=p$ the system \eqref{eq:Phi-piece-wise-ab} has the following solutions
\begin{subequations}
\begin{align}
\label{eq:same-p-1}&b=\pm\frac{\pi}{2},\qquad a=\mp\frac{\pi}{2}\vartheta_cp\cos(au_c),\\
\label{eq:same-p-2}&a=k\pi,\qquad\sin b=(-1)^{k+1}\frac{pk}{2\vartheta_cu_c},
\end{align}
\end{subequations}
where $k\in\mathbb{Z}$. Taking into account that $\vartheta_cu_c\ll1$ one obtains from \eqref{eq:same-p-2} only the trivial solution $a=0$ and $b=k\pi$. Substituting the obtained solutions into Hamiltonian \eqref{eq:Energy} results that the trivial solution corresponds to the energy maximums and the solutions \eqref{eq:same-p-1} minimize the energy. Therefore for the case of equal polarities one obtains the solution \eqref{eq:Phi-sol-appr}.

For the case of opposite polarities $p_1=-p_2=p$ the system \eqref{eq:Phi-piece-wise-ab} results
%the system \eqref{eq:Phi-piece-wise-ab} has the following solutions has the following solutions
\begin{subequations}\label{eq:oppos-sol}
\begin{align}
\label{eq:oppos-p-1}&a=\frac{\pi}{2u_c}(2k+1),\quad \cos b=(-1)^k(2k+1)\frac{p}{u_cv_c},\\
\label{eq:oppos-p-2}&b=k\pi,\quad\sin(au_c)=(-1)^{k}a\frac{2p}{\pi\vartheta_c}.
\end{align}
\end{subequations}
Due to the condition $\vartheta_cu_c\ll1$ the system \eqref{eq:oppos-sol} results in the trivial solutions $a=0$ and $b=k\pi$, and consequently $\Phi=0$ or $\Phi=\pi$. Analysis of the energy functional shows that for $p=1$ the solution $\Phi=\pi$ and for $p=-1$ the solution $\Phi=0$ correspond to the energy minimum.
%---------------------------------------------
%\bibliography{soliton}
%\bibliography{bibliograpy}
%
%---------------------------------------------
\end{document}